\documentclass[aps,prb,twocolumn,superscriptaddress,showpacs]{revtex4}

\usepackage{graphicx}
\usepackage{dcolumn}
\usepackage{bm}

\begin{document}

\title{Low-temperature thermal conductivity of Dy$_2$Ti$_2$O$_7$ and Yb$_2$Ti$_2$O$_7$ single crystals}

\author{S. J. Li}
\affiliation{Hefei National Laboratory for Physical Sciences at Microscale, University of Science and Technology of China, Hefei, Anhui 230026, People's Republic of China}

\author{Z. Y. Zhao}
\email{zzhao20@utk.edu}

\affiliation{Hefei National Laboratory for Physical Sciences at Microscale, University of Science and Technology of China, Hefei, Anhui 230026, People's Republic of China}

\affiliation{Department of Physics and Astronomy, University of Tennessee, Knoxville, Tennessee 37996-1200, USA}

\author{C. Fan}
\affiliation{Hefei National Laboratory for Physical Sciences at Microscale, University of Science and Technology of China, Hefei, Anhui 230026, People's Republic of China}

\author{B. Tong}
\affiliation{Hefei National Laboratory for Physical Sciences at Microscale, University of Science and Technology of China, Hefei, Anhui 230026, People's Republic of China}

\author{F. B. Zhang}
\affiliation{Hefei National Laboratory for Physical Sciences at Microscale, University of Science and Technology of China, Hefei, Anhui 230026, People's Republic of China}

\author{J. Shi}
\affiliation{Department of Physics, University of Science and Technology of China, Hefei, Anhui 230026, People's Republic of China}

\author{J. C. Wu}
\affiliation{Hefei National Laboratory for Physical Sciences at Microscale, University of Science and Technology of China, Hefei, Anhui 230026, People's Republic of China}

\author{X. G. Liu}
\affiliation{Hefei National Laboratory for Physical Sciences at Microscale, University of Science and Technology of China, Hefei, Anhui 230026, People's Republic of China}

\author{H. D. Zhou}
\affiliation{Department of Physics and Astronomy, University of
Tennessee, Knoxville, Tennessee 37996-1200, USA}

\affiliation{National High Magnetic Field Laboratory, Florida
State University, Tallahassee, Florida 32306-4005, USA}

\author{X. Zhao}
\affiliation{School of Physical Sciences, University of Science and Technology of China, Hefei, Anhui 230026, People's Republic of China}

\author{X. F. Sun}
\email{xfsun@ustc.edu.cn}

\affiliation{Hefei National Laboratory for Physical Sciences at Microscale, University of Science and Technology of China, Hefei, Anhui 230026, People's Republic of China}

\affiliation{Key Laboratory of Strongly-Coupled Quantum Matter Physics, Chinese Academy of Sciences, Hefei, Anhui 230026, People's Republic of China}

\affiliation{Collaborative Innovation Center of Advanced Microstructures, Nanjing, Jiangsu 210093, People's Republic of China}

\date{\today}

\begin{abstract}
We study the low-temperature thermal conductivity ($\kappa$) of Dy$_2$Ti$_2$O$_7$ and Yb$_2$Ti$_2$O$_7$ single crystals in magnetic fields up to 14 T along the [111], [100] and [110] directions. The main experimental findings for Dy$_2$Ti$_2$O$_7$ are: (i) the low-$T$ $\kappa(H)$ isotherms exhibit not only the step-like decreases at the low-field ($<$ 2 T) magnetic transitions but also obvious field dependencies in high fields ($>$ 7 T); (ii) at $T \le$ 0.5 K, the $\kappa(H)$ curves show anisotropic irreversibility in low fields, that is, the $\kappa(H)$ hysteresis locates at the first-order transition with $H \parallel$ [100] and [110], while it locates between two successive transitions with $H \parallel$ [111]; (iii) the $\kappa$ in the hysteresis loops for $H \parallel$ [100] and [110] show an extremely slow relaxation with the time constant of $\sim$ 1000 min. The main experimental findings for Yb$_2$Ti$_2$O$_7$ are: (i) the zero-field $\kappa(T)$ show a kink-like decrease at the first-order transition ($\sim$ 200 mK) with decreasing temperature; (ii) the low-$T$ $\kappa(H)$ isotherms show a decrease in low field and a large enhancement in high fields; (iii) the low-$T$ $\kappa(H)$ curves show a sharp minimum at 0.5 T for $H \parallel$ [110] and [111]. The roles of monopole excitations, field-induced transitions, spin fluctuations and magnetoelastic coupling are discussed.

\end{abstract}


\maketitle

\section{INTRODUCTION}

Rare-earth titanates $R_2$Ti$_2$O$_7$ ($R =$ rare earth) have attracted extensive research interests due to their exotic magnetism.\cite{Review} These materials have a pyrochlore crystal structure with the space group $Fd\bar{3}m$, in which the magnetic rare-earth ions form a network of corner-sharing tetrahedra and are prone to a high degree of geometric frustration. However, these materials are greatly sensitive to weak perturbations (e.g. single-ion anisotropy, dipolar interaction or quantum fluctuations) beyond the nearest-neighboring exchange, which results in unconventional low-temperature magnetic and thermodynamic properties. One famous phenomenon is the classical spin-ice state in Ho$_2$Ti$_2$O$_7$ and Dy$_2$Ti$_2$O$_7$ (with effective ferromagnetic exchange and Ising anisotropy).\cite{spin ice-1, spin ice-2, spin ice-3, spin ice-4, spin ice-5} At low temperatures, their moments have an Ising anisotropy due to the strong crystal field with the local easy axis along the [111] axis. The ground states have a macroscopically degenerate ``2-in, 2-out'' spin configuration in each tetrahedron. An interesting finding in the spin-ice materials is that the magnetic excitations could be emergent magnetic monopoles.\cite{monopole-1, monopole-2, monopole-3, monopole-4, monopole-5, monopole-6, monopole-7, monopole-8, monopole-9, monopole-10, monopole-11, monopole-12} Once the flipping of a spin occurs, a local ``3-in, 1-out'' or ``3-out, 1-in'' spin configuration forms, which is equivalent to yielding two opposite magnetic monopoles in the adjacent tetrahedra.\cite{monopole-1}

For Dy$_2$Ti$_2$O$_7$, the external magnetic field along different directions can easily break the degenerate spin-ice state and induce various magnetic states.\cite{Fukazawa} For example, with increasing magnetic field along the [100] axis, the system enters a Q = 0 state at $\mu_0H >$ 0.5 T.\cite{spin ice-1, Fennell} This state has a single spin configuration chosen from the degenerate spin-ice ground state, with all the spins on each tetrahedron having a component along the field direction and actually forming a long-range order. When the field is along the [110] direction, the system enters a Q = X state at $\mu_0H >$ 0.4 T.\cite{Hiroi, Fennell, Yoshida1, Melko} The spin system is separated into two sets of chains parallel ($\alpha$) and perpendicular ($\beta$) to the field, with long-range ferromagnetic order and short-range antiferromagnetic order, respectively.\cite{Hiroi, Fennell, Yoshida1, Melko} A more complicated case occurs when applying field along the [111] direction. There are two successive transitions from the spin-ice state to the kagom\'e-ice state and then to the fully-polarized state at $\sim$ 0.3 and 0.9 T, respectively.\cite{magnetic transition-1, magnetic transition-2, magnetic transition-3, magnetic transition-4, magnetic transition-5} Another notable characteristics of the spin-ice materials is the very slow spin dynamics at low temperatures, demonstrated by the dc magnetization, ac susceptibility, and specific-heat measurements, etc.\cite{monopole-5, monopole-8, monopole-11, Snyder, Matsuhira, Yaraskavitch, Pomaranski, Paulsen} Due to the slow spin dynamics, these field-induced magnetic transitions show significant irreversibility, which has been probed by the magnetization and neutron scattering.\cite{monopole-8, magnetic transition-1, Fennell}

Yb$_2$Ti$_2$O$_7$ has been proposed to be a good candidate for the quantum version of spin ice.\cite{QSI-1, QSI-2, QSI-3} The crystal-field structure of the Yb$^{3+}$ ion is a ground-state Kramers doublet well-separated from the first excited doublet. The $g$-factor has large planar component, $g_{\perp}$ = 4.18, compared to the [111] component $g_{\parallel}$ = 1.77.\cite{g-factor} The net interaction between the neighboring Yb$^{3+}$ ions is ferromagnetic with a Curie-Weiss temperature of $\sim$ 0.65 K.\cite{g-factor, CW} For these factors, this material can be described as an effective pseudospin-1/2 quantum-spin ice, with strong transverse quantum fluctuations of magnetic dipoles.\cite{QSI-1, QSI-2, QSI-3} The nature of ground state of Yb$_2$Ti$_2$O$_7$ is still rather controversial. The specific-heat measurements have revealed a sharp peak at $\sim$ 200 mK, signifying a first-order transition.\cite{first order-1, first order-2, first order-3, first order-4} However, the neutron scattering measurements have indicated that the low-$T$ state is weakly ferromagnetic, accompanied with short-range and dynamic spin correlations.\cite{neutron-1} Furthermore, the neutron scattering above 200 mK has revealed the pinch-point structure, indicating the presence of a spin-liquid phase with spin-ice correlations.\cite{first order-2} The excitations of magnetic monopoles are expected in this ``high"-$T$ quantum spin-ice state.\cite{first order-3, Applegate, Pan-1}

Low-temperature heat transport is a powerful tool to probe the properties of elementary excitations\cite{Berman, Brenig, Hess, Sologubenko1, Yamashita, Taillefer, Sun_LSCO, Sun_YBCO} and the field-induced magnetic transitions.\cite{CuGeO3, Sologubenko2, Sologubenko3, Sun_DTN, Zhao_GFO, Zhao_NCO, Zhang_GdErTO} In principle, the magnetic monopoles, as the elementary excitations in the spin-ice state, can also contribute to the heat transport by acting as either heat carriers or phonon scatterers. In a pioneer work by Klemke {\it et al.},\cite{Klemke} it was concluded that the thermal conductivity ($\kappa$) of Dy$_2$Ti$_2$O$_7$ is purely phononic and the field dependence of $\kappa$ is attributed to phonon scattering by monopoles. In contrast, Kolland {\it et al.} explained their $\kappa$ data of Dy$_2$Ti$_2$O$_7$ in terms of the magnetic monopoles making a large contribution to the $\kappa$ in zero field.\cite{Kolland1, Kolland2, Schareffe1, Schareffe2} Toews {\it et al.} measured the low-$T$ $\kappa$ of another spin-ice material, Ho$_2$Ti$_2$O$_7$, and analyzed the data with the considerations that the magnetic monopoles act as heat carriers as well as phonon scatterers.\cite{Toews} Recently, we have also studied the low-$T$ thermal conductivity of Dy$_2$Ti$_2$O$_7$ single crystals with $H \parallel$ [111] and the heat current parallel or perpendicular to the (111) plane.\cite{Fan_DTO} The zero-field $\kappa(T)$ was proved to be a purely phononic heat transport,\cite{Fan_DTO} consistent with the conclusion by Klemke {\it et al.}\cite{Klemke} Furthermore, an irreversible $\kappa(H)$ behavior was observed in the field region where the magnetization shows a nearly reversible plateau. A picture of the pinning effect of magnetic monopoles by the weak disorders was proposed to explain this novel phenomenon.\cite{Fan_DTO} It seems that an accurate understanding on the low-$T$ heat transport of Dy$_2$Ti$_2$O$_7$ has not been achieved.

A very recent work on Yb$_2$Ti$_2$O$_7$ has found a suppression of thermal conductivity in low fields.\cite{Tokiwa} It was explained as a large monopole heat transport at $T >$ 200 mK. In physics, the quantum monopoles can be more promising than the classical monopoles for transporting heat because they are dispersive while the classical ones are dispersionless. A coherent propagating of monopoles was also indicated by the terahertz spectroscopy and microwave cavity techniques.\cite{Pan-2} However, more detailed studies on the temperature and field dependencies of thermal conductivity are desperately needed for establishing a complete understanding of the heat transport properties.

In this work, we study in details the low-$T$ thermal conductivity of Dy$_2$Ti$_2$O$_7$ and Yb$_2$Ti$_2$O$_7$ single crystals in magnetic fields up to 14 T along the [111], [100] and [110] directions. It is found that most results of Dy$_2$Ti$_2$O$_7$ are not supportive to a sizeable heat transport of monopoles, while some results of Yb$_2$Ti$_2$O$_7$ point to a considerable quantum monopole transport at $T >$ 200 mK. The $\kappa$ at very low temperatures is mainly the phonon transport, but the spin fluctuations in Yb$_2$Ti$_2$O$_7$ strongly scatter phonons. In both materials, the field-induced magnetic transitions can affect the phonon heat transport strongly. Moreover, the magnetoelastic coupling are probably also playing a role. All these indicate rather complicated mechanisms of the heat transport in these materials.

\section{EXPERIMENTS}

\begin{figure}
\includegraphics[clip,width=8.5cm]{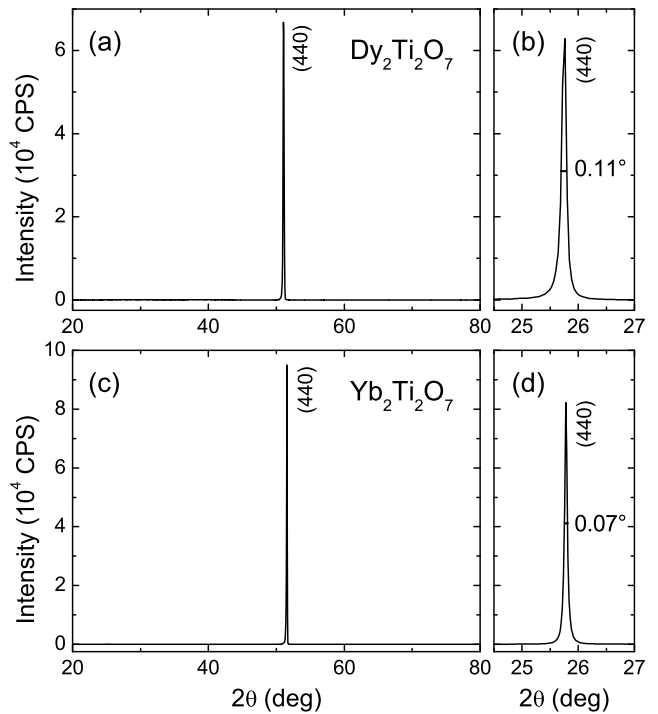}
\caption{X-ray diffraction pattern of (110) plane (a and c) and the rocking curve of (440) peak (b and d) for two pieces of Dy$_2$Ti$_2$O$_7$ and Yb$_2$Ti$_2$O$_7$ single crystals, which were orientated by using the X-ray Laue photographs. The full width at half maximum (FWHM) of the rocking curves are shown in panels (b) and (d).}
\end{figure}

High-quality Dy$_2$Ti$_2$O$_7$ and Yb$_2$Ti$_2$O$_7$ single crystals were grown using a floating-zone technique.\cite{Li_Growth} The basic low-temperature properties of our crystals, including the magnetic susceptibility (down to 2 K) and specific heat (down to 0.4 K), were checked and found to be in good consistency with most of the reported data from the literatures. Some of these data were already shown in Ref. \onlinecite{Li_Growth}. To exhibit the crystallinity of samples, some representative XRD data on the single crystals are shown in Fig. 1. The narrow widths of the rocking curve of the (440) Bragg peak (FWHM $\approx 0.11^\circ$ and $\approx 0.07^\circ$ for these two samples), demonstrates that the crystals have good crystallinity.

For the $\kappa$ measurements, the long-bar shaped samples were cut from the as-grown crystals along the [111], [100] or [110] axes (the uncertainty is smaller than 1$^\circ$) after orientation by using back-reflection X-ray Laue photographs. The $\kappa$ was measured using a ``one heater, two thermometers" technique in a $^3$He refrigerator at 300 mK $< T <$ 30 K and a $^3$He-$^4$He dilution refrigerator at 50 mK $< T <$ 1 K, equipped with a 14 T magnet.\cite{Sun_DTN, Zhao_GFO, Zhao_NCO, Zhang_GdErTO, Fan_DTO, Zhao_IPA} All the measurements were done with both the magnetic field and the heat current along the longest dimension of the samples, which can minimize the demagnetization effects.\cite{Fan_DTO}

The thermal conductivities of Dy$_2$Ti$_2$O$_7$ along the [111], [100], and [110] axes were measured at low temperatures down to 0.3 K and in magnetic fields up to 14 T on three samples with the sizes of 3.1$\times$0.71$\times$0.15 mm$^3$, 3.08$\times$0.66$\times$0.15 mm$^3$ and 2.86$\times$0.66$\times$0.15 mm$^3$, respectively. In this work, the magnetic-field dependencies of $\kappa$ for Dy$_2$Ti$_2$O$_7$ were measured in two different processes. The first one is the standard steady-state technique with the field changing step by step.\cite{Sun_DTN, Zhao_GFO, Zhao_NCO, Zhang_GdErTO, Fan_DTO} That is, the measurements were done by the following steps: (i) change field slowly to a particular value and keep it stable; (ii) after the sample temperature is stabilized, apply a heat power at the free end of sample; (iii) wait some time (typically for several minutes at very low temperatures) until the temperature gradient on the sample is nearly stabilized (judged from the time dependencies of two RuO$_2$ thermometers on the samples; a relative change of less than 1 \% within a time window of 2 or 3 minutes is the usual criterion); (iv) record the temperatures of two thermometers and obtain the temperature gradient. Using this process, the $\kappa(H)$ isotherms for three samples were measured at different temperatures and the irreversibilities were also probed with the measurements done in changing field (from zero after zero-field cooling) up and down in the step mode. The speed of changing field is slow enough to avoid observable heating effect from the eddy current. More exactly, the field ramping rate is not a constant, and it must be very slow at $\mu_0H <$ 0.1 T (0.001--0.01 T/min) and could be quicker at $\mu_0H >$ 0.1 T (0.01--0.05 T/min). Note that this process is suitable for most of solid materials. However, it is known that Dy$_2$Ti$_2$O$_7$ has very slow spin dynamics at low temperatures,\cite{Matsuhira} which may cause some relaxation phenomenon in the temperature gradient measurements,\cite{Kolland1, Schareffe1} if the magnetic excitations are involved in the heat transport behaviors. For this reason, the $\kappa$ measurement done with the sweeping-field mode could not be very meaningful because the thermal relaxation and spin relaxation must be strongly mixed.

To probe the time relaxation of the $\kappa$, the measurements were also done in a second process, which is different from the first one in the last step. Namely, after the field and sample temperature are stabilized, apply a suitable heat power at the free end of sample and wait for about 30 minutes;  then, record the time dependence of temperature gradient on the sample. Since the relaxation was found to be extremely slow at some particular fields, it is not possible to finish a relaxation measurement at different fields in a up and down loop. The relaxations were measured for three different samples at very low temperatures. In this case, the $\kappa$ at a certain field was measured separately in the field-increasing and field-decreasing processes, respectively. In the former case, the sample was cooled to the measurement temperature (for example, 0.36 K) in zero field and then field was changed slowly to the particular values, and then the $\kappa$ was measured in each fields. In the latter case, the sample was first cooled to 0.36 K in zero field; after that the field was swept slowly to 1--2 T, which are well above the irreversible region as we show below, then decreased slowly to the target field.

The thermal conductivities of Yb$_2$Ti$_2$O$_7$ along the [100], [110], and [111] axes were measured at low temperatures down to 50 mK and in magnetic fields up to 14 T on three samples with dimensions of $2.07 \times 0.64 \times 0.12$ mm$^3$, $2.87 \times 0.57 \times 0.14$ mm$^3$ and $2.53 \times 0.60 \times 0.14$ mm$^3$, respectively. The measurements of Yb$_2$Ti$_2$O$_7$ single crystals are much simpler, since there is no observable relaxation effect.

\section{RESULTS}

\subsection{Zero-field $\kappa(T)$ of Dy$_2$Ti$_2$O$_7$}

\begin{figure}
\includegraphics[clip,width=7.5cm]{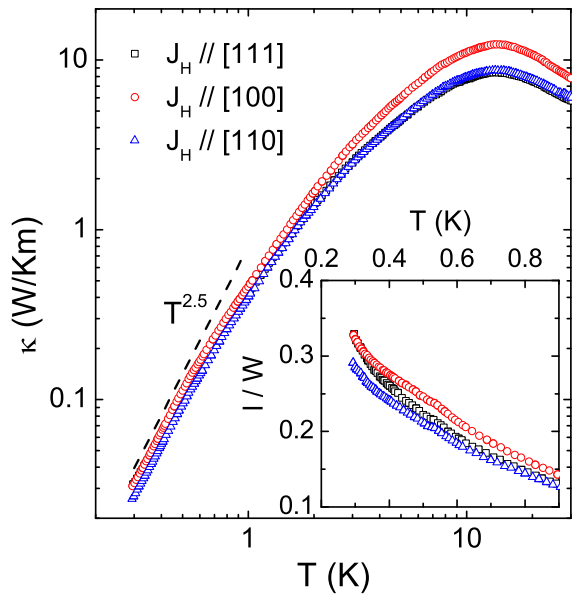}
\caption{(Color online) Temperature dependencies of the thermal conductivities of Dy$_2$Ti$_2$O$_7$ in zero field for heat current along the [111], [100] or [110] axis. The dashed line shows the $T^{2.5}$ temperature dependence. The inset shows the temperature dependencies of the phonon mean free path $l$ divided by the averaged sample width $W$ for these samples.}
\end{figure}

Figure 2 shows the temperature dependencies of $\kappa$ for Dy$_2$Ti$_2$O$_7$ along three directions and in zero field. Apparently, the heat transport is nearly isotropic and shows a simple phonon transport. The phonon peaks are located at $\sim$ 15 K and a rough $T^{2.5}$ dependence of $\kappa$ is visible at subkelvin temperatures, which is however weaker than the standard $T^3$ behavior of phonon thermal conductivity at the boundary scattering limit.\cite{Berman} Similar result has been obtained in the earlier works and the magnetic scattering on phonons was discussed to be important at temperatures below 10 K.\cite{Klemke, Kolland1, Kolland2, Schareffe1, Schareffe2, Fan_DTO} The mean free path of phonons at low temperatures are calculated in the usual way.\cite{Fan_DTO, Zhao_GFO, Sun_Comment} The phononic thermal conductivity can be expressed by the kinetic formula $\kappa_{ph} = Cv_pl/3$,\cite{Berman} where $C = \beta T^3$ is the phonon specific heat at low temperatures, $v_p$ is the average velocity and $l$ is the mean free path of phonons. The $\beta$ value of Dy$_2$Ti$_2$O$_7$ crystals, obtained from our specific-heat data,\cite{Li_Growth} is $1.14 \times 10^{-3}$ J/K$^4$mol (which is in good consistency with the result from other group\cite{magnetic transition-2}). The inset to Fig. 2 shows the ratios $l/W$ for three samples, where $W$ is the averaged sample width.\cite{Fan_DTO, Zhao_GFO, Sun_Comment} It is clear that the ratios increase quickly at very low temperatures and are expected to approach 1 at $T <$ 0.3 K. This calculated result is compatible with the temperature dependence of $\kappa$, which is a bit weaker than $T^3$. Note that if there were other type of heat carriers (e.g. magnetic monopoles) that makes a large contribution to the $\kappa$, the actual phonon mean free path would be much smaller than those in Fig. 2, which is apparently not very reasonable.

\subsection{$\kappa(H)$ of Dy$_2$Ti$_2$O$_7$}

The low-$T$ thermal conductivities of Dy$_2$Ti$_2$O$_7$ had been studied in magnetic fields up to only 7 T and the main finding was that the $\kappa$ can be strongly suppressed by rather weak fields ($<$ 1 T).\cite{Klemke, Fan_DTO, Kolland1, Kolland2, Schareffe1, Schareffe2} There are two different explanations for the field-induced suppression of $\kappa$ in Dy$_2$Ti$_2$O$_7$. Namely, it is due to either the suppression of a monopole heat transport or the field-enhanced magnetic scattering on phonons.\cite{Fan_DTO, Kolland1, Kolland2, Schareffe1, Schareffe2} Note that the $\kappa(T)$ data shown in Fig. 2 are not supportive for a sizeable monopole heat transport in zero field. In passing, a recent work on another spin-ice material, Ho$_2$Ti$_2$O$_7$, proposed that magnetic monopoles are playing a dual role in the heat transport; that is, they can both transport heat and scatter phonons.\cite{Toews} In that work, the contribution of the monopole transport to the total thermal conductivity was estimated to be about 10 \%,\cite{Toews} much smaller than that proposed for Dy$_2$Ti$_2$O$_7$.\cite{Kolland1, Kolland2, Schareffe1, Schareffe2}

\begin{figure}
\includegraphics[clip,width=8.5cm]{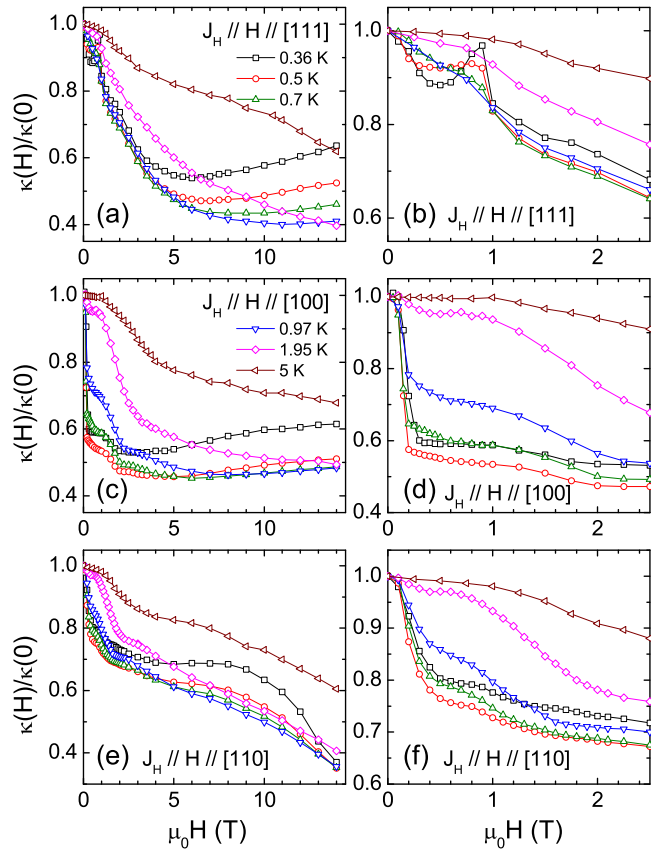}
\caption{(Color online) Magnetic-field dependencies of thermal conductivities of Dy$_2$Ti$_2$O$_7$: (a,c,e) data in fields up to 14 T and along three directions; (b,d,f) the zoom-in of low-field data. In this figure, the magnetic field is increased step by step after cooling the samples in zero field. The directions of the magnetic field and heat current ($J_H$) are the same.}
\end{figure}

In present work, the magnetic-field dependencies of $\kappa$ are measured up to 14 T, as shown in Fig. 3. The low-field behaviors are almost the same as those results in earlier studies. For $H \parallel$ [111], at very low temperatures, the $\kappa$ display two anomalies in low fields: at 0.36 K, the $\kappa(H)$ curve shows a quick decrease and a peak-like transition at about 0.25 and 0.9 T, respectively, which are in good correspondence with the critical fields of the subsequential transitions from the low-field spin-ice state to the kagom\'e-ice state then to the saturated state (``3-in, 1-out" or ``3-out, 1-in").\cite{magnetic transition-1, magnetic transition-2, magnetic transition-3, magnetic transition-4} At 0.5 K, the $\kappa(H)$ curve behaves like two step decreases at these two transitions. The sharpness of the transitions is quickly smeared out upon increasing temperature. For $H \parallel$ [100], the 0.36-K $\kappa(H)$ curve shows a steep and broad step-like transitions at 0--0.2 T and 1--2 T, respectively. Upon increasing temperature, the first transition gradually becomes weaker while the second one is enlarged and shifts to higher fields. At 5 K, the first step is completely gone and the second one evolves into a broad transition at 1--4 T. The first transition is in good correspondence with the critical field of a long-range-order transition,\cite{Fennell} in which the external magnetic field chooses one of the degenerate ground states with all the spins having the components along the field. For $H \parallel$ [110], the low-field $\kappa(H)$ behave rather similarly to the case of $H \parallel$ [100]. For example, at 0.36 K the $\kappa(H)$ curve also shows two step-like decreases at 0--0.25 and 0.8--1.5 T. The first step-like change of $\kappa$ at 0.25 T is in good correspondence with a first-order transition from the low-field spin-ice state to the spin-chain state,\cite{Hiroi, Fennell, Yoshida1, Melko} in which the $\alpha$-chains form a long-range ferromagnetic order and the $\beta$-chains form a short-range antiferromagnetic order. Upon increasing temperature, the first transition gradually becomes weaker, while the second one becomes larger and shifts to higher fields.

The drastic change of the heat transport at the magnetic phase transition has been widely observed in many magnetic materials and is usually related to the evolution of magnetic excitations.\cite{CuGeO3, Sologubenko2, Sologubenko3, Sun_DTN, Zhao_GFO, Zhao_NCO, Zhang_GdErTO} However, the particular features of $\kappa(H)$ at the magnetic transitions can be significantly different from each other, which depend on both the natures of magnetic transitions and the roles of magnetic excitations in the heat transport (as heat carriers or phonon scatterers). In the spin-ice state of Dy$_2$Ti$_2$O$_7$, the monopoles are the only one type of magnetic excitations. In this regard, it was naively assumed that the quick decreases of $\kappa$ at the low-field magnetic transitions are caused by the suppression of monopole transport.\cite{Kolland1, Kolland2, Schareffe1, Schareffe2} This scenario will be further analyzed in the following sections.

A remarkable finding is that these $\kappa(H)$ curves still exhibit clear or strong field dependencies when the field is above 7 T, which have not been achieved in those earlier works.\cite{Klemke, Fan_DTO, Kolland1, Kolland2, Schareffe1, Schareffe2} At subkelvin temperatures, the $\kappa(H)$ for $H \parallel$ [111] and [100] show a broad-valley-like feature; that is, after arriving a minimum at several tesla, the $\kappa$ gradually increases in high magnetic fields. In particular, at 0.36 K, the $\kappa$ at 14 T $\parallel$ [100] is already larger than that at 0.5 T, where the thermal conductivity was taken for the phononic background in a recent work.\cite{Kolland2} In addition, the $\kappa(H)$ for $H \parallel$ [111] and [100] are not yet saturated in 14 T.

It is notable that the high-field $\kappa(H)$ behave rather differently for $H \parallel$ [110]. The $\kappa(H)$ show weak field dependence at several tesla, particularly at 0.36 K, which is also consistent with the earlier reported data in field up to 7 T.\cite{Kolland2} However, at low temperatures the $\kappa$ show another strong decrease at $\mu_0H >$ 7 T and there is no sign that the $\kappa$ can be recovered even at 14 T. Since this phenomenon does not appear for fields along other directions, a high-field polarization effect is not likely the reason. For now, there are no other experimental results for Dy$_2$Ti$_2$O$_7$ in [110] field higher than 7 T, so it is unknown whether the strong suppression of $\kappa$ at $\mu_0H >$ 7 T ($\parallel$ [110]) is related to some magnetic transitions. For clarifying, high-field investigations by using other measurements are necessary in the near future.

The paramagnetic scattering effect related to the crystal-field levels of Dy$^{3+}$ ions was proposed to play an important role in the field dependencies of $\kappa$.\cite{Klemke, Fan_DTO} It is known that the Dy$^{3+}$ ions have a degenerate doublet of the lowest crystal-field level. In magnetic field, a Zeeman splitting of the doublet can produce resonant scattering on phonons and give both a low-field suppression and a high-field recovery of $\kappa$, as some other magnetic materials have shown.\cite{Sun_PLCO, Sun_GBCO, Zhao_NCO, Li_NGSO} However, since the paramagnetic scattering is a qualitatively isotropic effect, the obvious anisotropic and complicated high-field behaviors in Fig. 3 should not be solely relevant to this mechanism.

\subsection{Irreversibilities of $\kappa(H)$ of Dy$_2$Ti$_2$O$_7$}

\begin{figure}
\includegraphics[clip,width=8.5cm]{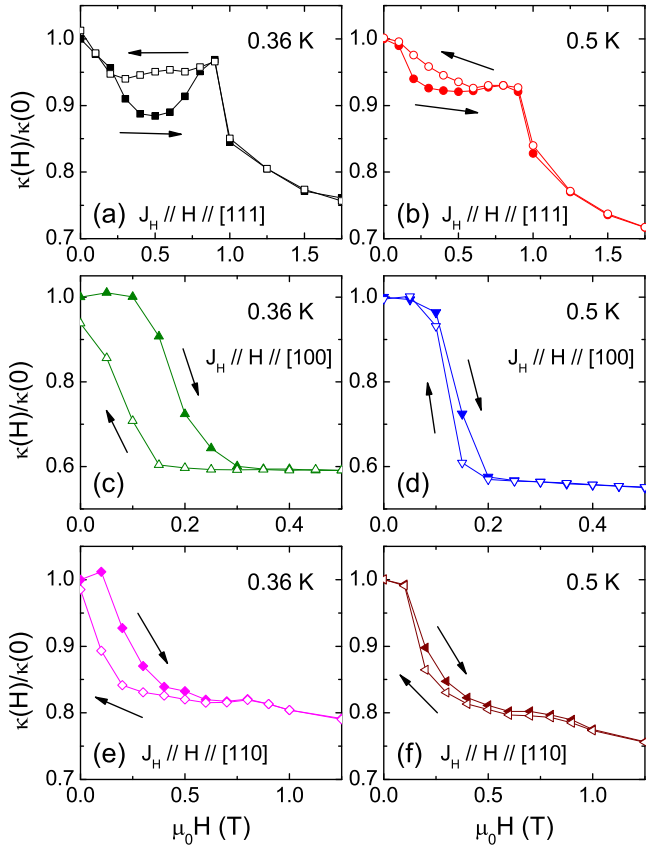}
\caption{(Color online) The low-$T$ $\kappa(H)$ loops with magnetic field sweeping up and down along the [111], [100], or [110] directions. The data shown with solid symbols are measured in ascending field after the sample is cooled in zero field, while the open symbols show the data with descending field, as also indicated by the arrows. These data were measured in the normal way, that is, the waiting time for the temperature stabilization is in a scale of ten minutes. Note that the irreversibilities are not visible at $T >$ 0.5 K (not shown here).}
\end{figure}

It is found that the magnetic-field dependencies of $\kappa$ display irreversible behavior at very low temperatures for all three field directions, as shown in Fig. 4. These data reproduce our earlier results for $H \parallel$ [111] and are similar to those reported by other group.\cite{Fan_DTO, Kolland1, Kolland2, Schareffe1, Schareffe2} The reversibility weakens quickly with increasing temperature and are hardly visible at $T > $ 0.5 K. Note that this is a bit different from some earlier results that clear hysteresis was observed at 0.6 K.\cite{Kolland1, Kolland2, Schareffe1, Schareffe2} The reason is that those data were taken in a sweeping-field mode.

There are significant differences in the irreversibility between the $\kappa(H)$ data with $H \parallel$ [111] and the data taken with $H \parallel$ [100] and [110]. First, the hysteresis of $\kappa(H)$ with $H \parallel$ [111] is located mainly in a field region between the two field-induced magnetic transitions, while those with $H \parallel$ [100] and [110] are located at the magnetic transitions. Actually, the irreversibility for $H \parallel$ [100] and [110] seems to be a common ``supercooling" phenomenon of a first-order phase transition. Note that the ultra-low-$T$ neutron scattering also demonstrated clear hysteresis of the field dependencies of Bragg scattering intensity for $H \parallel$ [100] and [110].\cite{Fennell} Second, the irreversibility for $H \parallel$ [111] appears in the kagom\'e-ice state, in which the magnetization curves do not show any hysteresis;\cite{monopole-7, magnetic transition-1, Fan_DTO, Kolland1, Kolland2} whereas those for $H \parallel$ [100] and [110] have good correspondence with the magnetization hysteresis.\cite{Kolland1, Kolland2, Schareffe1, Schareffe2} Third, the magnitudes of $\kappa$ are larger for decreasing fields than those for increasing field with $H \parallel$ [111], while the opposite behavior is observed with $H \parallel$ [100] and [110]. All these clearly indicate that the origin of the irreversibility with $H \parallel$ [111] is different from others.

\subsection{Relaxation phenomenon of $\kappa$ of Dy$_2$Ti$_2$O$_7$}

\begin{figure}
\includegraphics[clip,width=8.5cm]{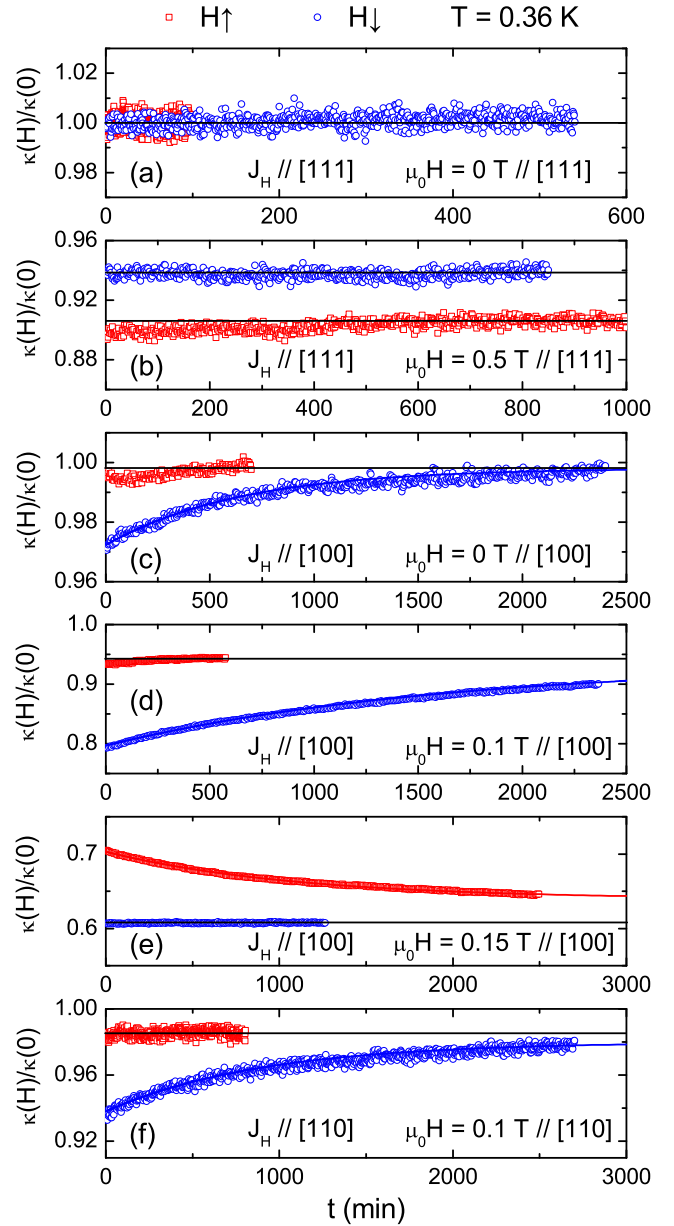}
\caption{(Color online) Time dependencies of thermal conductivities for field along the [111], [100], or [110] directions and at 0.36 K. The details for these measurements are described in the experimental section. The open squares show the data measured in ascending field after the sample is cooled in zero field, and the open circles with descending field from high field, where no relaxation and irreversibility are observed. The horizontal lines are guides for eyes while the curves are fittings using formula (\ref{E}). All the $\kappa(0)$ are taken the values from the normal measurements, as shown in Figs. 3 and 4.}
\end{figure}

It is known that the thermal conductivity of Dy$_2$Ti$_2$O$_7$ shows not only irreversible behaviors but also a time relaxation.\cite{Kolland1, Kolland2, Schareffe1, Fan_DTO} In present work, the relaxation effect is carefully studied and the representative data are shown in Fig. 5. Note that at 0.36 K, the relaxation of $\kappa$ can be very strong and very slow, which prevents us from getting the equilibrium state in the available time window of our $^3$He refrigerator. However, the relaxation effect is not visible in high magnetic fields or at $T \ge$ 0.5 K, where the $\kappa(H)$ hysteresis in Fig. 4 disappears.

Among three field directions, the relaxation of $\kappa$ in $H \parallel$ [111] is the weakest. As shown in Figs. 5(a) and 5(b), in either zero field or 0.5 T (the field showing largest hysteresis), the magnitude of relaxation is very small. Typically, after waiting for about 30 min (the starting point of the recorded data in Fig. 5), the measured $\kappa$ change in a scale of only 1--2 \%, which is nearly the same as the error of this measurement. This result confirms our earlier work that the relaxation effect in $H \parallel$ [111] is not significant.\cite{Fan_DTO} Although one cannot completely rule out the possibility of a relaxation with very long time scale (e.g. much longer than the time window of this measurement), the magnitude of the $\kappa$ changing with time for $H \parallel$ [111] is significantly smaller than those for other field directions. A bit larger relaxation of $\kappa$ for $H \parallel$ [111] could be observed in a different measurement process, in which the magnetic field was continuously swept.\cite{Schareffe1}

The relaxation of $\kappa$ is the most significant for $H \parallel$ [100]. As shown in Fig. 5(c), although there is no sizeable relaxation effect after cooling in zero field, decreasing field to zero from high field (1 T or higher) results in a slow relaxation. This is very different from the case of [111] field shown in Fig. 5(a). In addition, in non-zero fields, the relaxation could be very strong and is related to the history of applying field. For example, increasing field from 0 to 0.1 T does not cause significant relaxation, but decreasing field from high value to 0.1 T leads to a very large relaxation; the $\kappa$ can be changed about 10 \% and is not saturated after a very long time of about 2400 min, as shown in Fig. 5(d). A contrary feature is observed at 0.15 T; as shown in Fig. 5(e), decreasing field from high value to 0.15 T does not cause significant relaxation, but increasing field from 0 to 0.15 T leads to a very slow relaxation. Furthermore, the $\kappa$ slowly relaxes towards larger values in the case of decreasing field to 0.1 T, while it slowly relaxes towards smaller values if increasing field to 0.15 T. Such differences demonstrate that the hysteresis of $\kappa(H)$ is directly related to the relaxation effect, as discussed in the following text.

The time dependence of $\kappa$ can be fitted by a single exponential function,
\begin{equation}
\frac{\kappa(H)}{\kappa(0)} = a + be^{-t/\tau}, \label{E}
\end{equation}
where $\tau$ is the characteristic time constant, and $a$ and $b$ are time independent parameters. The time constant $\tau$ could be as large as 1100--1600 min.

The situation for $H \parallel$ [110] is very similar to that for $H \parallel$ [100], but the relaxation is a bit weaker. For example, the time constants $\tau$ are about 900 min and the $\kappa$ changes about 5 \% with decreasing field to 0.1 T, as shown in Fig. 5(f).

The slow spin dynamics of the spin ice has been probed by other experiments on Dy$_2$Ti$_2$O$_7$, such as the dc magnetization, ac susceptibility, and specific heat.\cite{monopole-5, monopole-8, monopole-11, Snyder, Matsuhira, Yaraskavitch, Pomaranski, Paulsen} It has been found that the spin relaxation time in the spin-ice state showed an exponential increase at low temperatures and is greater than 10$^4$ s at $T <$ 0.45 K. The current explanations are mainly in terms of diffusive motion of monopoles. After cooling in zero field to a low but finite temperature, the density of monopoles decreases gradually with time. This could induce a slow relaxation with large time scale. However, it is notable in the present $\kappa$ measurements that cooling in zero field does not lead to a significant relaxation of $\kappa$. This means that the monopole dynamics does not have close correlation with the heat transport of Dy$_2$Ti$_2$O$_7$. On the other hand, the magnetization measurements with the field-quench or pulse-field processes have revealed the monopole dynamics having a time scale of only several minutes.\cite{monopole-9, monopole-11} This is in contrast with the extremely slow relaxation of $\kappa$ in magnetic field. More importantly, if the relaxation of $\kappa$ in finite fields, as shown in Figs. 5(d--f), were mainly caused by the monopole dynamics (either transporting heat or scattering phonons), one would expect opposite trends of $\kappa(t)$ for increasing or decreasing to the same field. Again, none of the data supports this expectation. Therefore, the relaxation of $\kappa$ cannot be well explained based on the assumption that the monopoles play an important role in the heat transport properties. Furthermore, the anisotropic relaxation of $\kappa$ for different field directions indicates that the slow spin dynamics associated with the field-induced transitions for $H \parallel$ [100] and [110] are strongly coupled with phonons, whereas that for $H \parallel$ [111] is not.

\begin{figure}
\includegraphics[clip,width=6.0cm]{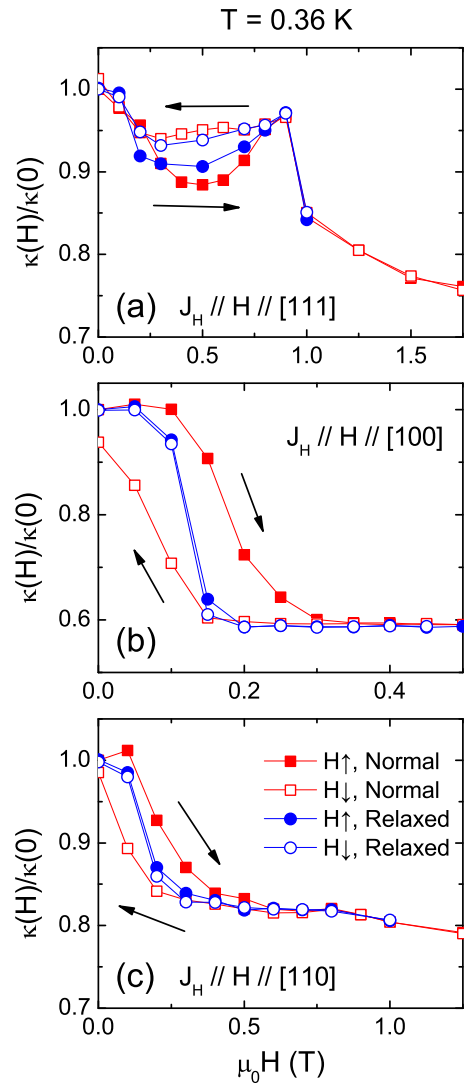}
\caption{(Color online) The 0.36-K $\kappa(H)$ loops with the magnetic field sweeping up and down along the [111], [100], or [110] directions. The data shown with squares are measured in the normal way, that is, the waiting time for the temperature stabilization is in a scale of several or ten minutes. The circles show the data extrapolated at $t \rightarrow \infty$ from the time-relaxation measurements. The arrows indicate the direction of changing field.}
\end{figure}

The relaxation fittings using formula (\ref{E}) can get not only the time constant but also the extrapolated values of $\kappa(H)/\kappa(0)$ at $t \rightarrow \infty$. Then, the magnetic-field dependencies of $\kappa$ in the expected ``equilibrium or completely relaxed state" can be obtained. Figure 6 shows the comparison of the $\kappa(H)$ between the data of normal measurements and the data after complete relaxation for all three field directions. The main result is that the irreversibility in the [111] field changes only slightly and still clearly exists after long-time relaxation, but the irreversibilities in the [100] and [110] fields almost disappear for the data extrapolated at $t \rightarrow \infty$. This demonstrates that the irreversibilities of $\kappa(H)$ in the [100] and [110] fields are related to the extremely slow spin dynamics in Dy$_2$Ti$_2$O$_7$, while that in the [111] field has a different origin. As discussed in our previous work, a pinning effect of magnetic monopoles by crystal disorders can explain the particular irreversibility with $H \parallel$ [111].\cite{Fan_DTO} It is also notable that in this case the irreversibility of $\kappa(H)$ is actually not large, which is consistent with the previous conclusion on the minor role of monopoles in the heat transport.

\subsection{Analysis on the Dy$_2$Ti$_2$O$_7$ data assuming the monopole transport}

\begin{figure}
\includegraphics[clip,width=8.5cm]{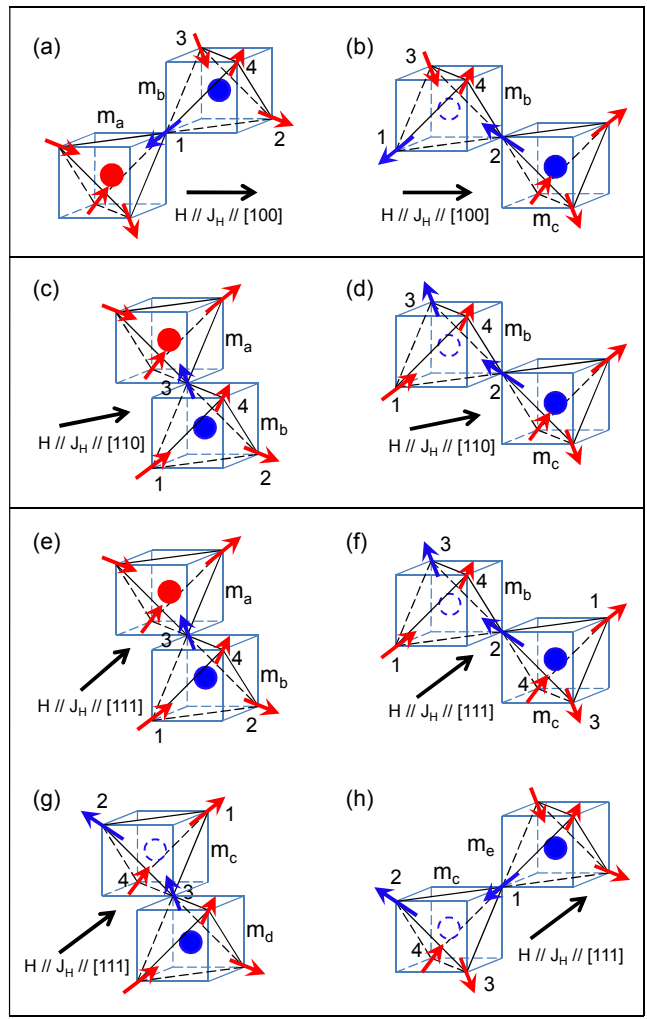}
\caption{(Color online) Schematic plot for the motions of magnetic monopoles under magnetic fields along three directions. Note that only those motions that can contribute to heat transport in the [100] (a,b), [110] (c,d) and [111] (e--h) directions ($J_H \parallel H$) are shown. Red and blue dots represent magnetic monopoles with opposite polarities. Blue arrows represent the flipped spins, associated with a monopole movement from the dashed-dot position to the solid-dot position.}
\end{figure}

The present experimental results have already met with some difficulties in explaining the data when assuming large monopole heat transport. A quantitative analysis on the low-field decrease of $\kappa$ is useful to make further clarification. Figure 7 shows the schematic picture of monopole motions that can contribute to the heat transport along the magnetic-field directions.

The changes of energy gap between the ground state and excited state and the density of monopoles are dependent on the field direction.\cite{Tokiwa, Castelnovo} When the field is applied along the [100] direction, the ground state and excited state split into three and two levels, respectively.\cite{Castelnovo} Because the split levels are rather complicated, a simplification should be carried out. According to the calculation of Castelnovo {\it et al.},\cite{Castelnovo} the density of monopoles decays in an approximately exponential speed because the energy gap increases. Here, only the hopping from the lowest level of ground state to the lowest level of excited state is considered, whose gap is $\Delta=\Delta_0+(2/\sqrt{3})g\mu_BH$ and $\Delta_0 =$ 4.35 K is the energy gap in zero field.\cite{LDC, Castelnovo} Figure 7(a) presents this process, in which the spin 1 flips and a pair of monopoles $m_a$ and $m_b$ is created. $m_b$ has three choices of spin flip to recover to the ground state. If the spin 1 flips, the two monopoles annihilate, and there is no monopole movement. If one of the other two equivalent spins 2 or 4 flips, $m_b$ moves to the next tetrahedron and becomes $m_c$, as shown by Fig. 7(b). This movement has a positive component along the heat current and makes contribution to thermal conductivity. After this process involving the heat transport, the energy of the two tetrahedra is lifted by $E=(2/\sqrt{3})g\mu_BH$. On one hand, increasing field leads to larger $E$. As a result, this process becomes more and more difficult and the movement of monopoles are restricted. On the other hand, the energy gap $\Delta$ for monopoles also increases and fewer monopoles can be excited.

In fields along the [110] direction, both the ground state and the excited state split into three levels. With the same simplification, the energy gap between the lowest levels of ground state and excited state is $\Delta = \Delta_0$; that is, the gap for monopoles is field independent for $H \parallel$ [110]. This means the density of monopoles also does not vary with field. Figure 7(c) shows a pair of monopoles, $m_a$ and $m_b$, created by this process. Apart from the annihilation caused by the spin 3, the inequivalent spins 2 and 4 provide other two recovery choices for $m_b$. If the spin 4 flips, the monopole propagates along the direction perpendicular to heat current, and this process makes no contribution to heat transport. If the spin 2 flips, as shown by Fig. 7(d), $m_b$ moves to the next tetrahedron and becomes $m_c$, with a positive component along the heat current. This process accompanied by the heat transport leads to an energy lift of $E=(2\sqrt{6}/3)g\mu_BH$. This energy lift increases with field, and therefore the monopole movement is suppressed.

In the case of $H \parallel$ [111], the propagation of monopoles along the [111] direction is the most complicated. The ground state and excited state split into two and four levels in the [111] field, respectively. The excitation gap of monopoles $m_a$ and $m_b$, as shown in Fig. 7(e), is $\Delta=\Delta_0-(2/3)g\mu_BH$, which results in an increase of monopole density. For the movement of $m_b$, two effective flips caused by the spins 2 and 4 are equivalent and lift the energy by $E_1=(2/3)g\mu_BH$. However, $m_c$, which is generated in this process shown by Fig. 7(f), does not leave the original (111) plane. Further effective flips have to be considered. First, if the next flipping occurs with the spin 3, as shown in Fig. 7(g), $m_c$ moves within the (111) plane and becomes $m_d$. Though energy drops by $E_2=(2/3)g\mu_BH$ in this process, no heat transport along the [111] direction occurs. Second, if the spin 1 flips, it can be seen in Fig. 7(h) that $m_c$ goes across the (111) plane and contributes to heat transport. $m_c$ turns into $m_e$ and energy is lifted by $E_3=2g\mu_BH$. Thus, two-step flips are needed to transporting heat, and the total energy lift $E=E_1+E_3$ also increases with field.

\begin{figure}
\includegraphics[clip,width=8.5cm]{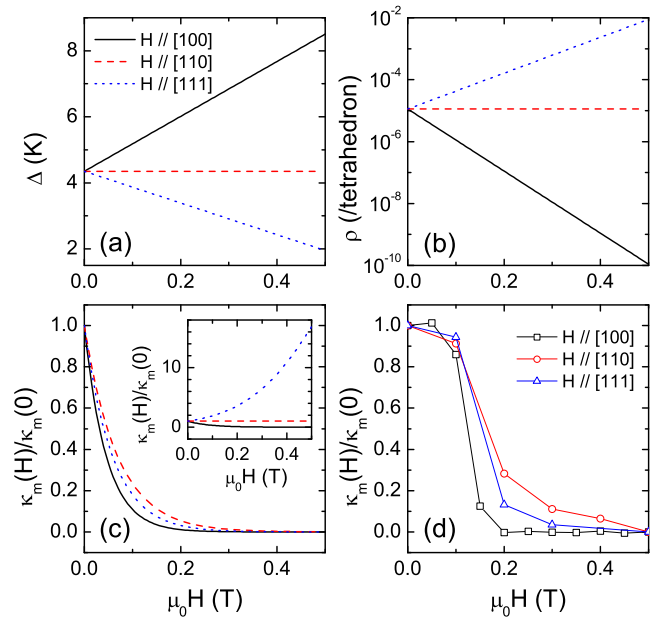}
\caption{(Color online) (a) Magnetic-field dependencies of the energy gap of monopole excitations in Dy$_2$Ti$_2$O$_7$. (b) Magnetic-field dependencies of the monopole density. (c) Calculated monopole thermal conductivity using the kinetic formula after considering the field dependencies of $\Delta$, $\rho$ and velocity of monopole. The inset shows the calculation using the same formula, but the field dependencies of velocity are not considered. (d) The monopole thermal conductivity extracted from the experimental data in Fig. 6, assuming the decrease of raw-data $\kappa$ from 0 to 0.5 T is just a decrease of $\kappa_m$. All the calculations and experimental data are taken at 0.36 K.}
\end{figure}

The above discussions indicate that magnetic field affects both the excitation and the propagation of monopoles. The former refers to the monopole energy $\Delta$ and the density $\rho$, while the latter refers to velocity $v$. As discussed above, the monopole energy $\Delta$ with $H \parallel$ [100], [110] and [111] are $\Delta_0+(2/\sqrt{3})g\mu_BH$, $\Delta_0$ and $\Delta_0-(2/3)g\mu_BH$, respectively, as shown in Fig. 8(a). For the monopole density, only free monopoles are concerned. According to the Debye-H\"{u}ckle theory,\cite{Castelnovo} the monopole density is determined by the energy gap,
\begin{equation}\label{eq:eps}
\rho=\frac{2\exp{(\Delta/k_BT)}}{1+2\exp{(\Delta/k_BT)}}.
\end{equation}
This formula gives the field dependence of the monopole density, as shown in Fig. 8(b). The specific heat of monopoles can be obtained by\cite{Castelnovo}
\begin{equation}\label{eq:eps}
C_m=N\frac{\partial}{\partial T}(\rho\Delta),
\end{equation}
where $N$ represents the number of tetrahedra per unit volume. Since the positive and negative monopoles would annihilate when they meet, the mean free path of monopoles is assumed to be the averaged distance of monopoles with the same polarity. Thus, it can be written as
\begin{equation}\label{eq:eps}
l_m=(N\frac{\rho}{2})^{-1/3}.
\end{equation}

If the velocity of monopoles is independent of field, according to the kinetic formula $\kappa_m = C_mv_ml_m/3$, $\kappa_m(H)/\kappa_m(0)$ at 0.36 K can be calculated and are shown in the inset to Fig. 8(c). This calculation shows that the $\kappa_m$ increases strongly with $H \parallel$ [111] and is independent of field with $H \parallel$ [110], which are completely different from the experimental observations. In fact, the velocity of monopole should be sensitive to the magnetic field and its direction, as discussed above. Currently, the velocity of monopoles can not be calculated accurately. If only the relative change is concerned, based on the additional energy lift $E$ caused by the magnetic field, we simply assume that the monopole velocity decays exponentially with increasing $E$; that is,
\begin{equation}\label{eq:eps}
v_m = v_{m0}\exp{(-\frac{E}{\alpha k_BT})},
\end{equation}
where $v_{m0}$ is the monopole velocity at zero field and $\alpha$ is an adjustable parameter. $E$ are given by $E=(2/\sqrt{3})g\mu_BH$, $E=(2\sqrt{6}/3)g\mu_BH$ and $E=(8/3)g\mu_BH$ for $H \parallel$ [100], [110] and [111], respectively. With this additional assumption, $\kappa_m(H)/\kappa_m(0)$ at 0.36 K are calculated and shown in Fig. 8(c).

For comparison, Fig. 8(d) shows the low-field ``$\kappa_m$ data" at 0.36 K extracted from the experimental results for $H \parallel$ [100], [110] and [111]. Here, the ``relaxed" $\kappa(H)$ data from Fig. 6 are used, since the above calculations do not take into account the relaxation effect. It is assumed that the decrease of $\kappa$ from 0 to 0.5 T is simply due to the suppression of $\kappa_m$. The raw-data of $\kappa$ at 0.5 T is taken as the background from the phonon transport. It seems that there is a rather good consistency between the calculations and the experimental data.

Furthermore, from the assumed $\kappa_m$ data shown in Fig. 8(d), the absolute values of $\kappa_m(0)$ along the [100], [110] and [111] directions are $2.39\times10^{-2}$, $8.85\times10^{-3}$ and $6.01\times10^{-3}$ W/Km, respectively. This anisotropy can be explained as the difference in monopole ratio, $n$, that conducts heat in different directions. Based on the above discussion, only certain spin flips contribute to transporting heat. If we assume the probabilities of spin flips in one tetrahedron are equal and omit flips leading to all-in or all-out state, the probability for a monopole conducting heat is 2/3, 1/3 and 2/9 in the [100], [110] and [111] directions, respectively, which gives the monopole ratio, $n$, participating in transporting heat. From Eqs. (2), (3) and (4), the monopole density, specific heat and mean free path at 0.36 K are 1.13 $\times$ 10$^{-5}$ per tetrahedron, 176.88 J/Km$^3$ and 2.83 $\times$ 10$^{-8}$ m, respectively. Using the formula $\kappa_m = nC_mv_{m0}l_m/3$, the $v_{m0}$ in the [100], [110] and [111] directions can be calculated to be 2.12 $\times$ 10$^4$, 1.59 $\times$ 10$^4$ and 1.62 $\times$ 10$^4$ m/s, respectively. These values are unreasonably too large since monopoles are known to be dispersionless.\cite{Tokiwa}

Finally, we need to point out another experimental phenomenon that cannot be explained using the assumption of monopole heat transport. As found in our former work, for $H \parallel$ [111] the field dependencies of $\kappa$ along the field and perpendicular to it are nearly isotropic.\cite{Fan_DTO} As discussed above, these two perpendicular monopole transport are established with the processes shown by Figs. 7(e)--7(f)--7(h) and Figs. 7(e)--7(f)--7(g), respectively. In the former case, the process in Fig. 7(h) is suppressed most strongly by magnetic field. According to magnetic susceptibility result, the spin 1 in Fig. 7(e) is fixed along field direction at about 0.3 T,\cite{magnetic transition-1} which leads to the disappearance of $\kappa_m(H)$. In the latter case, magnetic field only suppresses the process in Fig. 7(f). At 0.3 T, although fixing spin 1 does not forbid the monopole transport directly, it reduces the degree of freedom of monopoles and forbids the subsequent monopole propagating after the flip of spin 1. Therefore, $\kappa_m(H)$ should have a kink-like decrease at 0.3 T instead of completely disappears. When field reaches 0.9 T, the spin 2 in Fig. 7(f) is also fixed\cite{magnetic transition-1} and the $\kappa_m$ is completely suppressed. Therefore, the $\kappa_m(H)$ along the [111] field and perpendicular to it are expected to present obvious anisotropy, which is not supported by the experimental results.\cite{Fan_DTO}

The analysis and calculations indicate that only some of the low-field results of $\kappa(H)$ can be qualitatively explained as the suppression of monopole thermal conductivity in magnetic fields. However, the quantitative analysis also points out that the monopoles are not likely making a large contribution to the heat transport. It is essentially coincided with some other experiments. For example, it has been found that the hop rate of monopoles is roughly proportional to the density of monopoles at low temperatures ($<$ 1.5 K).\cite{Castelnovo} Furthermore, a recent adiabatic susceptibility measurement revealed a Brownian characteristic of the monopole movement in the spin-ice state,\cite{monopole-12} with a much smaller diffusion constant than that proposed by some heat transport work.\cite{Kolland1} There should be more important factors that are responsible for the field dependencies of $\kappa$ in Dy$_2$Ti$_2$O$_7$.

\subsection{Zero-field $\kappa(T)$ of Yb$_2$Ti$_2$O$_7$}

\begin{figure}
\includegraphics[clip,width=7.0cm]{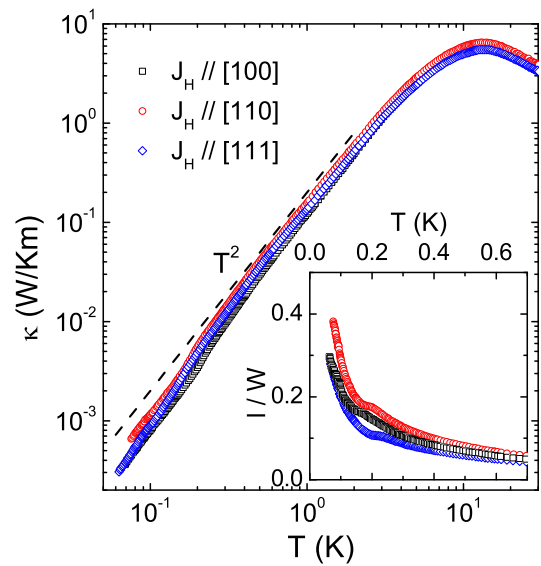}
\caption{(Color online) Temperature dependencies of the thermal conductivity along the [100], [110] and [111] axes of Yb$_2$Ti$_2$O$_7$ single crystals in zero field. The dashed line shows the $T^2$ dependence which the $\kappa$ follow between 200 mK and 2 K. The inset shows the temperature dependencies of the ratio of the phonon mean free path $l$ to the averaged sample width $W$. The $l$ is calculated assuming that the $\kappa$ is purely phononic.}
\end{figure}

Figure 9 shows the temperature dependence of $\kappa$ of Yb$_2$Ti$_2$O$_7$ single crystals in zero field. Similar to the case of Dy$_2$Ti$_2$O$_7$, the thermal conductivity is nearly isotropy for the heat current along the [100], [110] and [111] axes. The peak at 13 K is the typical feature of a phonon transport in insulators. Moreover, there are two notable features of these data. First, the temperature dependencies display kink-like anomalies at $\sim$ 200 mK where the slopes of $\kappa(T)$ changes strongly. This is likely related to the first-order transition of magnetism. Second, at subkelvin temperatures, the $\kappa$ do not show the $T^3$ dependence, which is expected for phonon thermal conductivity in the boundary scattering limit. Actually, the $\kappa$ nearly follow the $T^2$ dependence from about 2 K to 200 mK. This is consistent with a recent report.\cite{Tokiwa} Although the well-known $T^3$ ballistic behavior of phonons has been rarely observed in the transition-metal compounds, including the high-$T_c$ cuprates, the multiferroic manganites and the low-dimensional quantum magnets,\cite{Taillefer, Sun_YBCO, Sun_Nonuniversal, Sologubenko1, Sologubenko2, Yamashita, Zhao_GFO, Sun_DTN, Chen_MCCL} the $\kappa$ can usually exhibit the temperature dependence rather close to $T^3$. In Dy$_2$Ti$_2$O$_7$, as shown in Fig. 2, the $T^{2.5}$ dependence of $\kappa$ is observed at temperatures of several hundreds of millikelvins. In Yb$_2$Ti$_2$O$_7$, however, the $T^2$ dependence differs too much from the $T^3$ behavior. There are two possible reasons for this deviation from the boundary scattering limit. If there are sizeable scattering effects on phonons even at the lowest temperature, the phonon mean free path is significantly smaller that the sample size and $T^3$ behavior cannot be observed. Another possibility is the presence of the other term of heat conductivity, for example, from the magnetic excitations. In any case, the estimation of phonon mean free path can provide some useful information.

With the $\beta$ value (= 1.61 $\times 10^{-3}$ J/K$^4$mol) from the specific-heat data,\cite{Li_Growth} the phonon mean free path can be calculated assuming the $\kappa$ is purely phononic. The inset to Fig. 9 shows the temperature dependencies of the ratio $l/W$. It is found that the $l/W$ ratios are only about 0.3--0.4 at the lowest temperatures, indicating that the boundary scattering limit of phonons is not established and there are still some microscopic scattering of phonons. It should be noted that the above calculation is based on the assumption of a purely phononic heat transport. If there were a magnetic term of heat transport, the phonon mean free path would be even smaller. Nevertheless, it is clear that the phonon scattering is stronger in Yb$_2$Ti$_2$O$_7$, compared to Dy$_2$Ti$_2$O$_7$. It is understandable from the magnetic scattering since Yb$_2$Ti$_2$O$_7$ is known to have strong spin fluctuations.\cite{neutron-1}

All the zero-field $\kappa(T)$ curves show a kink-like decrease at 200 mK with lowering temperature. It might be that some term of thermal conductivity disappears when entering the low-$T$ state. This is likely a signature that the magnetic excitations (monopoles) in the quantum spin-ice state (at $T >$ 200 mK) can act as heat carriers.\cite{Tokiwa}

\subsection{$\kappa(H)$ of Yb$_2$Ti$_2$O$_7$}

\begin{figure}
\includegraphics[clip,width=8.5cm]{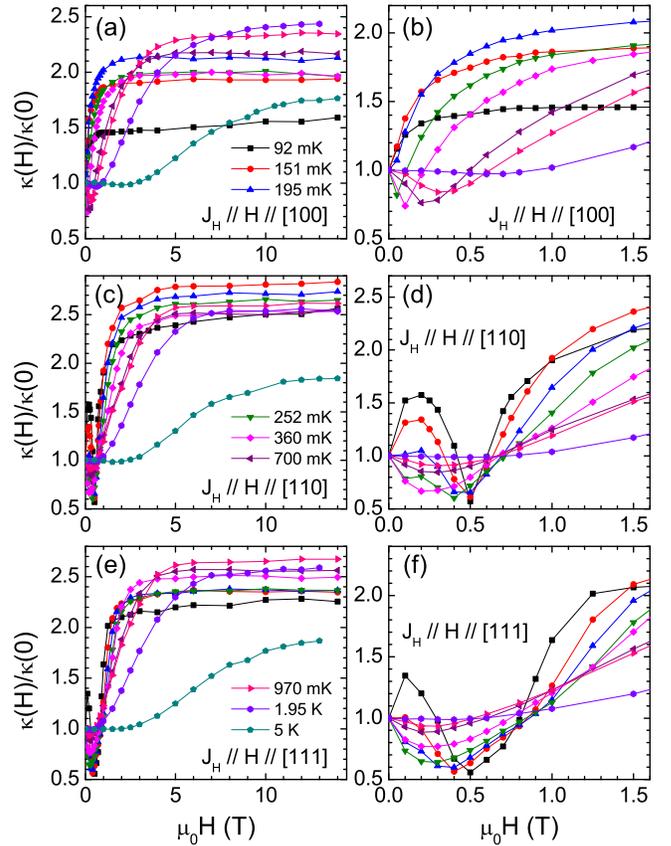}
\caption{(Color online) Magnetic-field dependencies of the thermal conductivity of Yb$_2$Ti$_2$O$_7$ crystals at low temperatures. Panels (a,c,e) show the data in fields up to 14 T and along three directions, while panels (b,d,f) show the zoom-in data of the low-field plots. Since the magnetic fields are applied along the direction of the heat current ($J_H$), the demagnetization effect is negligibly small for these long-bar shaped samples.}
\end{figure}

Figure 10 shows the magnetic-field dependencies of $\kappa$ at low temperatures for these Yb$_2$Ti$_2$O$_7$ single crystals. The data have two remarkable features. First, with increasing field, the $\kappa$ exhibits strong changes in low fields, followed by a quick increase, and finally becomes field independent in high fields. It naively indicates that there are magnetic scattering effects on phonons that can be suppressed in high fields, where the spins are polarized. This is consistent with the zero-field $\kappa(T)$ data and the small phonon mean free path. The strong spin fluctuations revealed by the neutron scattering is most likely the source of phonon scattering in zero field. Second, the low-field behaviors are somewhat different for different directions of magnetic field. For $H \parallel$ [100], the $\kappa$ increase monotonically with field when $T <$ 200 mK, but show a low-field valley when $T >$ 200 mK. The position of dip field shifts to higher field with increasing temperature. For $H \parallel$ [110] or [111], the $\kappa(H)$ isotherms at the lowest temperatures show a ``dip" at 0.5 T. In particular, the dip is very sharp for $H \parallel$ [110], reminiscent of a field-induced transition. In this regard, the absence the ``dip" for $H \parallel$ [100] indicates that either there is no field-induced transition or the transition field is very low (smaller than 0.1 T, which is the step size of our $\kappa(H)$ measurements).

As suggested in a recent work, the low-field decrease of $\kappa$ at $T >$ 200 mK may be due to the suppression of monopole heat transport in magnetic field.\cite{Tokiwa} However, it should be pointed out that the mechanism of monopole transport is far from well understood. The present $\kappa(H)$ data with field along three different directions show some peculiar phenomenon. To illustrate, at $T >$ 200 mK the low-field behavior of $\kappa$ are qualitatively different for different field directions. The position of dip field shifts to higher field with increasing temperature for $H \parallel$ [100], while it changes very weakly for $H \parallel$ [110] and [111]. This anisotropic behavior calls for a more precise theoretical description of the monopole transport.

Qualitatively, a low-field valley and a strong enhancement at high field in $\kappa(H)$ are also similar to the paramagnetic scattering effect.\cite{Berman, Sun_GBCO, Sun_PLCO, Zhao_NCO, Li_NGSO} It is known that the Yb$^{3+}$ ions have a Kramers doublet of the lowest crystal-field level. In magnetic fields, a Zeeman splitting of the doublet can produce resonant scattering on phonons and give both a low-field suppression and a high-field recovery of $\kappa$.\cite{Sun_PLCO, Sun_GBCO, Zhao_NCO, Li_NGSO} A simplified calculation has shown the qualitative behaviors of $\kappa(H)$ in the case of paramagnetic scattering; one important feature is that the position of the $\kappa(H)$ minimum shifts to higher field with increasing temperature.\cite{Sun_GBCO, Li_NGSO} Since the changes of low-field valley with increasing temperature are not in good consistency with this calculation, the paramagnetic scattering may not be important in Yb$_2$Ti$_2$O$_7$.

The low-field behavior of $\kappa$ at $T <$ 200 mK displays different characteristics from the data at $T >$ 200 mK. The low-$T$ magnetization measurements have indicated that the ground state is ferromagnetic with the spin easy-axis in the [100] direction, but the residual spin fluctuations are evidenced.\cite{Lhotel} These results indicated that the spins are easily polarized in a rather weak field. In this regard, the quick increase of $\kappa$ with field seems closely related to the spin polarization, which suppresses the spin fluctuations. In the field along the [110] direction, neutron scattering has revealed a field-induced transition at 0.5 T.\cite{neutron-1} With field lower than 0.5 T, there is significantly diffuse scattering intensity originated from the spin fluctuations. With increasing field, the diffuse scattering intensity is weakened and well-defined spin-wave excitations appear at 0.5 T, indicating a long-range ordered state (spin polarization).\cite{neutron-1} Since the spin waves are gapless at the critical field of 0.5 T, they can be well populated by the thermally excitations even at very low temperatures and strongly scatter phonons. This can explain the sharp 0.5T-``dip" of $\kappa(H)$ curves. Although there is no experimental investigations on this field-induced transition with field along other directions, the 0.5T-``dip" of $\kappa(H)$ curves with $H \parallel$ [111] may have the same origin.

\subsection{High-field $\kappa(T)$ of Yb$_2$Ti$_2$O$_7$}

\begin{figure}
\includegraphics[clip,width=8.5cm]{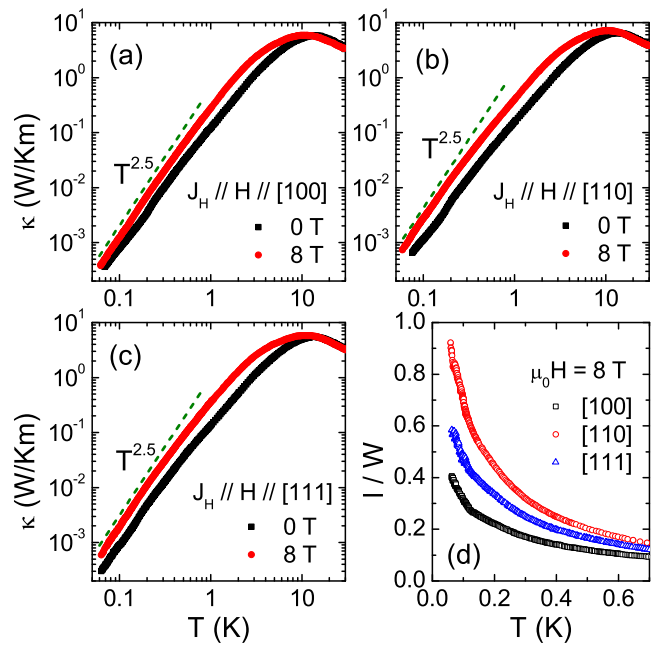}
\caption{(Color online) (a-c) Temperature dependencies of the thermal conductivity of Yb$_2$Ti$_2$O$_7$ crystals in zero field and in 8 T along different directions. The dashed lines indicate that the low-temperature data in 8 T follow the $T^{2.5}$ dependence. (d) Temperature dependencies of the ratio of the phonon mean free path $l$ to the averaged sample width $W$, calculated from the 8 T data.}
\end{figure}

In most cases, the magnetic scattering can be suppressed by strong magnetic field and the $\kappa$ would be independent of field in high fields and display a $T^3$ dependence at low temperatures. However, Yb$_2$Ti$_2$O$_7$ was found to be exceptional. Figure 11 shows the $\kappa(T)$ with 8 T along three different directions. It can be seen that these data do not exhibit the expected $T^3$ behavior even at $T <$ 100 mK; instead, the data follows a $T^{2.5}$ dependence at temperatures lower than 200-300 mK. The phonon mean free paths are calculated and are found to be smaller than the sample sizes, as shown in Fig. 11(d), which means that there are still some kind of microscopic scattering on phonons and the boundary scattering limit is not achieved. This temperature dependence of $\kappa$ in 8 T field is a peculiar phenomenon that is not easy to understand. Since high field can effectively suppress spin fluctuations, there should be some other factor damping the phonon transport.

\section{Discussion and summary}

Firstly, there is no strong evidence for the large heat transport of monopoles in the spin-ice compound Dy$_2$Ti$_2$O$_7$. Actually, most of the experimental data, including the zero-field $\kappa(T)$, the $\kappa(H)$ with different field directions, the high-field behaviors, and even the relaxation phenomenon, are more or less incompatible with the supposition of large monopole transport in zero field. The quantitative analysis using some phenomenological theory gives a qualitative description of the low-field decrease of $\kappa$ for $H \parallel J_H$, but it fails to explain the nearly isotropic behavior with $H \parallel$ [111] for $\kappa$ along the field and perpendicular to it.\cite{Fan_DTO} Furthermore, the calculations show that very high speeds ($>$ 10$^4$ m/s) of monopole propagating are required to describe the experimental data. This is unlikely since the monopole excitations are dispersionless and in principle they can only transport diffusively with a small averaged velocity. In the case of Yb$_2$Ti$_2$O$_7$, the kink-like feature of $\kappa(T)$ at 200 mK is a clear signature of considerable monopole transport at the quantum spin-ice state ($T >$ 200 mK). Since the quantum monopoles are dispersive, it is possible they have a highly mobile characteristic.

Secondly, judging from the $\kappa(T)$ data and the estimation of phonon mean free path, Dy$_2$Ti$_2$O$_7$ does not exhibit strong phonon scattering in zero field, while Yb$_2$Ti$_2$O$_7$ has rather strong scattering effect by the spin fluctuations. The high-field behaviors also demonstrate this difference. The main reason is that the monopole excitations are difficult to be thermally excited at very low temperatures due to the sizeable energy barrier ($\sim$ 4.35 K).

Thirdly, the field dependencies of $\kappa$ are much more complicated in Dy$_2$Ti$_2$O$_7$. One phenomenon is that the low-field transitions have strongest effect and cause step-like decreases of $\kappa$. Since both the qualitative and quantitative analysis based on the assumption of large monopole heat transport cannot describe the experimental results, some unknown phonon scattering processes are relevant. In Yb$_2$Ti$_2$O$_7$, the field dependencies are mainly determined by the suppression of spin fluctuations when the spins are polarized, and the field-induced transitions can cause a sharp dip-like anomaly in $\kappa(H)$.

Fourthly, for Dy$_2$Ti$_2$O$_7$, the $\kappa$ in low fields exhibit a strong relaxation effect and irreversibility. These phenomenon are not observed in the $\kappa$ measurements of Yb$_2$Ti$_2$O$_7$. This is mainly related to the extremely slow spin dynamics in the low-field states of Dy$_2$Ti$_2$O$_7$.

Fifthly, there are some peculiarity in the high-field heat transport of both materials. Dy$_2$Ti$_2$O$_7$ displays strong field dependencies of $\kappa$ in high field up to 14 T, which is quite anomalous considering the spins are polarized above $\sim$ 2 T.\cite{Fukazawa} Yb$_2$Ti$_2$O$_7$ shows field independent $\kappa$ in the polarized state. However, the $T^{2.5}$ behavior of high-field $\kappa(T)$ is not expectable if the magnetic scattering of phonons are completely suppressed. These indicate that besides the impacts associated with the low-field magnetic transitions and spin fluctuations, some other factors are also important.

One factor can affect the phonon transport is the magnetostriction and magnetoelastic coupling in these materials. For example, at the transition from the kagom\'e-ice phase to the saturated state with $H \parallel$ [111],\cite{magnetic transition-4} the ultrasound measurement revealed sharp anomalies of the sound velocity and sound attenuation.\cite{monopole-10} Furthermore, there is a low-field phenomenon of Dy$_2$Ti$_2$O$_7$ that is not mentioned in the above discussions. For $H \parallel$ [100] or [110] and at very low temperatures, the weaker and broader step-like decreases of $\kappa$ at 1--2 T could not be related to any magnetic transitions and should have a different origin. In fact, all the earlier experimental results of magnetization and neutron scattering have not found any magnetic transition or anomaly at 1--2 T for $H \parallel$ [100] or [110]. Instead, this phenomenon may be related to the magnetostriction, which exhibited a small anomaly at 2 T along the [100] direction.\cite{Kolland2} In this regard, the lower-field ($<$ 1 T) transitions can affect the $\kappa$ in the same way since they show stronger effect on crystal lattice.\cite{monopole-10} The high-field behavior of Yb$_2$Ti$_2$O$_7$ may be related to magnetoelastic coupling as well. In the spin-liquid Tb$_2$Ti$_2$O$_7$, a magnetoelastic mode is formed by the hybridization of the first excited crystal-field level and the transverse acoustic phonons.\cite{Fennell-2} As a result, the phonon transport of Tb$_2$Ti$_2$O$_7$ is so strongly damped that it behaves like a glassy state.\cite{Li_TTO} In Yb$_2$Ti$_2$O$_7$, the coupling of the sound wave to quantum fluctuations has been investigated by the sound-velocity and sound-attenuation measurement.\cite{Erfanifam} Some anomalies in temperature and field-dependent sound velocity and attenuation were observed and were attributed to the first-order phase transition. Moreover, these measurements (down to 20 mK) have found that both the sound velocity and sound attenuation change continuously with field up to 5 T, without showing any signature of saturation.\cite{Erfanifam} This indicated that the magnetoelastic coupling is significant even in high fields. The structural distortions caused by the magnetoelastic coupling would prevent the phonons from transporting ballistically.

In summary, the low-$T$ thermal conductivity of spin-ice Dy$_2$Ti$_2$O$_7$ and quantum spin-ice Yb$_2$Ti$_2$O$_7$ are studied. The detailed temperature and field dependencies of $\kappa$ indicate that the mechanisms of heat transport in these materials are rather complicated. The phonons are the main heat carriers and several factors including the magnetic-monopole excitations, the field-induced magnetic transitions, the phonon scattering by spin fluctuations, and probably the structural distortions caused by the magnetoelastic coupling are involved. One conclusion is that the monopoles in Dy$_2$Ti$_2$O$_7$ play a minor role, while the quantum monopoles in Yb$_2$Ti$_2$O$_7$ may make a sizable contribution in carrying heat. The comprehensive theories for quantitatively describing the experimental results are called for and would be very useful for understanding the exotic magnetisms of these materials. Further experimental investigations on the magnetoelastic coupling in high fields are also indispensable.

\begin{acknowledgements}

We thank R. P. Sinclair for the helpful assistant. This work was supported by the National Natural Science Foundation of China, the National Basic Research Program of China (Grant Nos. 2015CB921201 and 2011CBA00111), and the Fundamental Research Funds for the Central Universities (Program No. WK2030220014).

\end{acknowledgements}

\end{document}